\documentclass[aps,pra,showpacs,amsmath,amssymb,superscriptaddress,twocolumn]{revtex4}
\usepackage[colorlinks=true,urlcolor=blue,linkcolor=blue,citecolor=blue,breaklinks=true]{hyperref}
\usepackage{epsfig}
\usepackage{natbib}
\usepackage{color}
\usepackage{booktabs}
\usepackage{multirow}
\usepackage{exscale}
\usepackage{epsfig}
\usepackage{epstopdf}
\usepackage{bm}
\usepackage{graphicx,amssymb,dcolumn,bm,amsmath,subfigure}
\usepackage[sort&compress]{natbib}
\usepackage{amsfonts}
\usepackage[mathscr]{eucal}

\makeatletter
\def\fnum@figure#1{\figurename\nobreakspace\thefigure.}
\def\fnum@table#1{\tablename\nobreakspace\thetable
       \hspace{0.02em}}
\makeatother
\begin{document}

\preprint{Submitted to PRA}
\title{Multipole solitons in competing nonlinear media with an annular
potential}
\author{Liangwei Dong} \email{dlw\_0@163.com}
\address{Department of Physics, Zhejiang University of Science and Technology, Hangzhou, China, 310023}

\author{Mingjing Fan}
\address{Department of Physics, Shaanxi University of Science and Technology, Xi'an, 710021, China}
\author{Changming Huang}
\address{Department of Electronic Information and Physics, Changzhi University, Changzhi, Shanxi 046011, China}

\author{Boris A. Malomed}
\address{Instituto de Alta Investigacion, Universidad de Tarapaca, Casilla
7D, Arica, Chile\footnote{Sabbatical address}}

\begin{abstract}
We address the existence, stability, and propagation dynamics of multipole-mode solitons in cubic-quintic nonlinear media with an imprinted annular (ring-shaped) potential. The interplay of the competing nonlinearity with the potential enables the formation of a variety of solitons with complex structures, from dipole, quadrupole, and octupole solitons to
necklace complexes. The system maintains two branches of soliton families with opposite slopes of the power-vs.-propagation-constant curves. While the solitons' stability domain slowly shrinks with the increase of even number $n$ of lobes in the multipole patterns, it remains conspicuous even for $n>16$. The application of a phase torque gives rise to stable rotation of the soliton complexes, as demonstrated by means of analytical and numerical methods.
\end{abstract}

\maketitle

\vspace{2pc}

\section{Introduction}

\label{Sec1} Various types of self-bound states exist in diverse nonlinear
systems \cite{kivshar2003optical, yang20103, malomed2022}. Generic species
of such states include dipole \cite{krolikowski2000,ge_dipole_2011,yang2004}, quadrupole \cite{yang2004,zang2021}, and necklace solitons \cite{mihalache2003,dong2008,aleksic2020,li2020,dong2021}, as well as vortex
solitons carrying angular momentum \cite{zeng2020,wang2022,liu2023}. These
are fundamentally important modes in nonlinear optics, Bose-Einstein
condensates (BECs), plasmas, etc.

In addition to the ubiquitous cubic (Kerr) nonlinearity, a variety of
competing nonlinearities, aiming to arrest the two-dimensional (2D) critical
collapse, driven by the cubic self-focusing, and thus stabilize 2D solitons,
were proposed theoretically and observed experimentally \cite{malomed2022}.
Representative examples include quadratic-cubic [$\chi ^{(2)}:\chi ^{(3)}$]
\cite{PhysRevE.66.016613} and cubic-quintic [$\chi ^{(3)}:\chi ^{(5)}$,
alias CQ] \cite{Quiroga-Teixeiro97, PhysRevE.64.057601,michinel2004square}
optical media, where the focusing lower-order nonlinear term is necessary
for self-trapping of the localized states, while the higher-order defocusing
nonlinearity secures the arrest of the collapse. In this case, the nonlinear
index of refraction is negative at the peak (center) of the self-trapped
light beam, while remaining positive in the beam's wings. The dielectric
response of some materials used in nonlinear optics is accurately
approximated by the CQ combination of nonlinear terms \cite{Lawrence:98,Cid}. Spatial optical solitons were observed too in a setting featuring
quintic-septimal (focusing-defocusing) nonlinearities with negligible
third-order nonlinearity \cite{PhysRevA.90.063835}. Competing nonlinearities
have also been studied in plasma physics \cite{zakharov1971behavior}, as
well as in the context of Bose superfluids \cite{PhysRevLett.78.1215}.

Multipole solitons feature a periodic azimuthal structure consisting of an
even number $n$ of lobes (poles), which are distributed evenly on a ring. Adjacent lobes have opposite signs, hence $n$ cannot be odd (except for $n=1$, which corresponds to the axisymmetric ground-state mode). The simplest structured state is the dipole, which
corresponds to $n=2$. The thickness of the lobes is usually much smaller
than the effective radius of the ring. Though the diffraction of multipole
modes can be effectively stymied by the nonlinearity, the expansion in the
radial direction \cite{soljacic1998} or spiral rotation \cite{soljacic2000}
eventually destroys the patterns in the course of long propagation.

The concept of 2D multipole solitons was put forward in Refs \cite%
{desyatnikov2001,desyatnikov2002}, where stabilization of multipole
structures was achieved due to cross-phase-modulation coupling with a
nodeless field in a two-component system. The existence of such solitons
requires a focusing saturable nonlinearity. On the other hand, the expansion
of necklace-like beams in focusing Kerr media can be slowed down by the
angular momentum imprinted upon the pattern \cite{soljacic2001}. Another
method for suppressing the expansion of necklace beams was elaborated in
saturable systems with fractional diffraction \cite{dong2021}. Metastable
necklace beams were also predicted in settings with competing
quadratic-cubic \cite{kartashov2002} and cubic-quintic \cite%
{mihalache2003,mihalache2004} nonlinearities and observed in local \cite{grow2007} and nonlocal nonlinear media \cite{rotschild2006}.

Relevant alternatives for trapping stable multipole solitons are provided by
confinement, with radial modulation of the local refractive index inducing
an effective trapping potential \cite{yang2004,RevModPhys.83.247}. Although
local variation of the index is always small in comparison to its background
value, the corresponding potential may be sufficient to strongly affect
properties of nonlinear modes (in particular, to stabilize them).
Multipole-mode solitons can be supprted by different forms of photonic
lattices (potentials) and various types of optical waveguides \cite%
{yang2004-1,kartashov20062,dong2010,yang20052,rose2007,susanto2008,xia2013,wang20152, wang201822, dong20092,kartashov2005, hong20152,kartashov20093}. Particularly, multipole solitons may be made stable under appropriate
conditions in passive \cite{yang2004-1,yang20052,rose2007,susanto2008} and $%
\mathcal{PT}$-symmetric lattices \cite{wang20152,wang201822}, circular
waveguide arrays \cite{kartashov20093}, and axially symmetric Bessel
lattices \cite{kartashov2005,dong20092,hong20152}.

In addition to multipole solitons in optics, patterns with different values
of $n$ were also predicted in BECs with contact \cite{baizakov2006} and
dipole-dipole \cite{Raymond} interactions. In binary BECs, metastable
quantum droplets in the form of ring-shaped clusters were
constructed in systems modeled by the amended Gross-Pitaevskii equation with
the Lee-Huang-Yang (beyond-mean-field) correction \cite{kartashov20194}. Stable multipole quantum droplets were predicted in a weakly anharmonic potential \cite{dong20223}.

Though many efforts were put forth to suppress the instability of multipole
solitons induced by the repulsive force between adjacent poles, the search
for stable multipole solitons with large $n$ is still a challenging problem,
while most works were focused on the stability of dipoles and quadrupoles.
Very recently, Liu \emph{et al.} predicted that stable multipole solitons
can exist in CQ media modulated by a harmonic-Gaussian potential \cite%
{liu20232}. Yet, the stability region of necklace-like solitons quickly
shrinks with the increase of $n$, the solitons with $n\geq 8$ being unstable
in their entire existence domain.

In this work, we put forward a model admitting the self-trapping of stable
multipole solitons with large $n$. The CQ nonlinearity, combined with an
imprinted ring-shaped potential is shown to support two branches of
multipole families. The radius of the multipole complexes and the respective
value of $n$ can be controlled adjusting the radius of the potential ring.
The stability domain for the solitons shrinks \emph{very slowly} with the
growth of $n$, allowing the existence of stable modes with $n>16$, greatly
exceeding values of $n$ that were previously reported for stable multipole
solitons. Unstable lower-branch solitons can survive over a long propagation
distance, while unstable upper-branch ones are quickly destroyed.

The following presentation is arranged as follows. The model is formulated
in Section 2, and numerical results for static patterns, produced by a
systematic numerical analysis, are summarized in Section 3. Analytical and
numerical findings for rotating multipole solitons are reported in Section
4. The paper is concluded by Section 5.


\section{The model}

\label{Sec2}

We consider an optical beam propagating along the $z$ axis in a bulk medium
with the CQ nonlinearity and imprinted transverse modulation of the
refractive index, which induces the effective radial potential. This setting
is governed by the 2D nonlinear Schr\"{o}dinger equation (NLSE)\ for the
complex field amplitude $\Psi $, written here in the normalized form:
\begin{equation}
i\frac{\partial \Psi }{\partial z}=\left[ -\left( \frac{\partial ^{2}}{\partial x^{2}}+\frac{\partial ^{2}}{\partial y^{2}}\right) -pV(r)-|\Psi
|^{2}+|\Psi |^{4}\right] \Psi .  \label{Eq1}
\end{equation}%
The effective ring potential, with radius $r_{0}$, radial width $d$, and
depth $p$, is adopted in the form of
\begin{equation}
V(r)=\exp \left[ -(r-r_{0})^{2}/d^{2}\right] ,  \label{V}
\end{equation}%
where $r=\sqrt{x^{2}+y^{2}}$ [in terms of the similar Gross-Pitaevskii
equation (GPE), with $z$ replaced by time $t$, the potential is $-V(r)$].
Equation (\ref{Eq1}) conserves the net power (alias the energy flow) $P$,
angular momentum $M$, and Hamiltonian $H$:
\begin{gather}
P=\int \int |\psi |^{2}\text{d}x\text{d}y,  \notag \\
M=i\int \int \psi ^{\ast }\left( y\frac{\partial }{\partial x}-x\frac{\partial }{\partial y}\right) \psi \text{d}x\text{d}y,  \label{Eq3} \\
H=\int \int \left[ \left\vert \frac{\partial \psi }{\partial x}\right\vert
^{2}+\left\vert \frac{\partial \psi }{\partial y}\right\vert ^{2}+pV|\psi
|^{2}-\frac{1}{2}|\psi |^{4}+\frac{1}{3}|\psi |^{6}\right] \text{d}x\text{d}%
y,  \notag
\end{gather}%
where $\ast $ stands for the complex conjugate.

Stationary solutions of Eq.~(\ref{Eq1}) are looked for as
\begin{equation}
\Psi (x,y,z)=\psi (x,y)\exp (ibz),  \label{Psipsi}
\end{equation}%
where $\psi $ is the soliton's stationary profile, and $b$ is the
longitudinal propagation constant (in terms of the corresponding GPE, $-b$
is the chemical potential). Substituting ansatz (\ref{Psipsi}) in Eq.~(\ref%
{Eq1}), one obtains the stationary equation,
\begin{equation}
b\psi =\frac{\partial ^{2}\psi }{\partial x^{2}}+\frac{\partial ^{2}\psi }{%
\partial y^{2}}+pV\psi +|\psi |^{2}\psi -|\psi |^{4}\psi ,  \label{Eq2}
\end{equation}%
which can be solved numerically by means of the relaxation or
Newton-conjugate-gradient method \cite{yang20103}.

Soliton families are defined by propagation constant $b$, potential depth $p$, and geometric parameters $r_{0}$ and $d$. By means of scaling, we fix
\begin{equation}
r_{0}=2\pi ,  \label{r0}
\end{equation}%
and select $d=\sqrt{6}$ and $p=10$, which makes it possible to produce
generic results. Variation of the remaining free parameter $b$ produces
families of soliton states which are reported below. Other values of $d$ and
$r_{0}$ produce quite similar results.


\begin{figure}[tbph]
\centering
\includegraphics[width=0.48\textwidth]{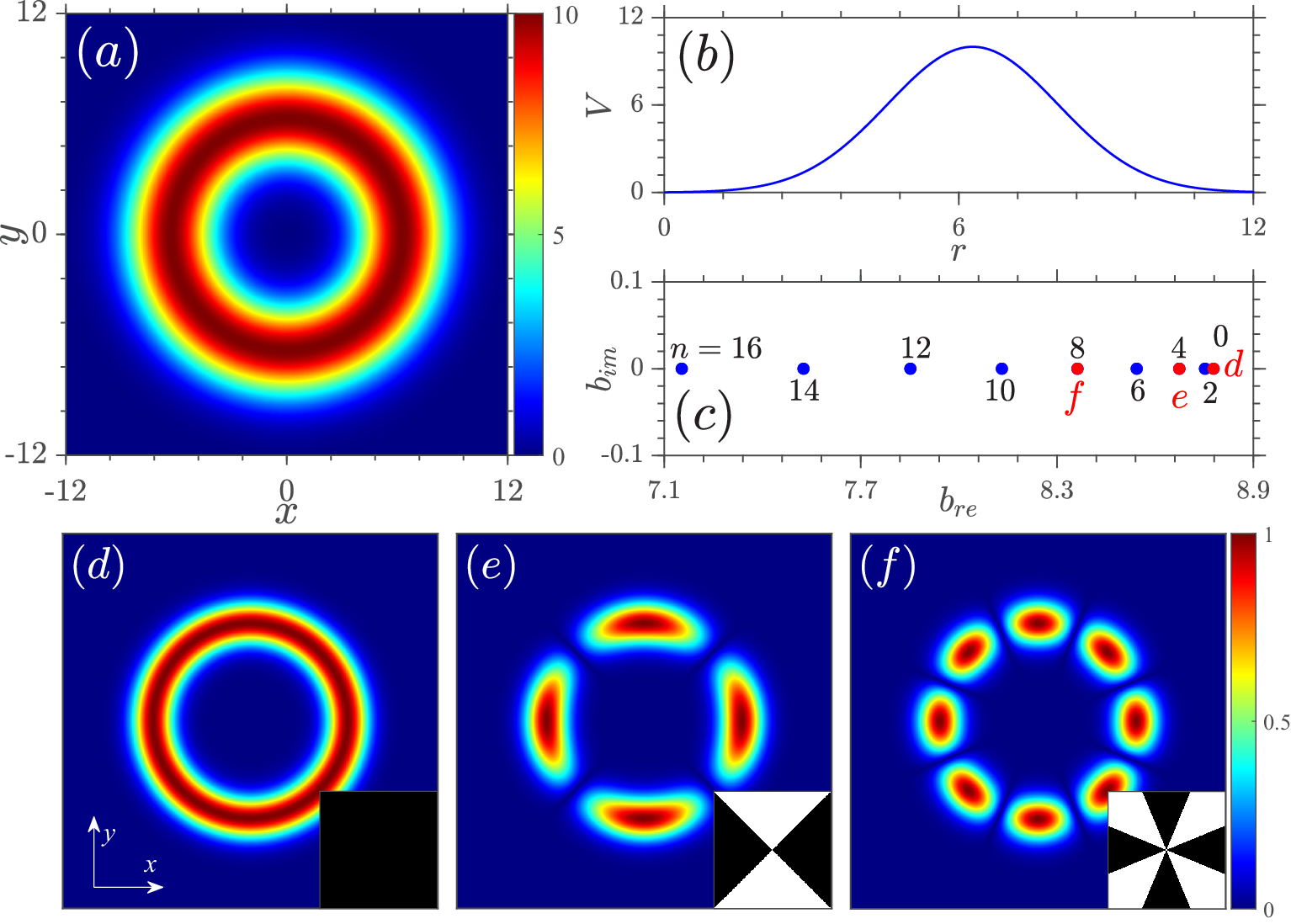}\vskip-0pc
\caption{The top view (a) and radial cross-section (b) of the ring potential
(\protect\ref{V}) with $p=10$, $r_{0}=2\protect\pi $, $d=\protect\sqrt{6}$.
(c) The respective spectrum of real eigenvalues produced by the linearized
equation (\protect\ref{Eq2}). The values of $n$ denote the pole number of the linear states from which multipole solitons bifurcate out. (d-f) Linear eigenstates of fundamental,
quadrupole and octupole modes corresponding to the eigenvalues marked in
(c). Insets in these panels and similar figures displayed below demonstrate
the phase structure of the eigenstates (alternation of values $0$ and $\protect\pi$).}
\label{fig1}
\end{figure}

The effective ring-shaped potential is displayed in Fig.~\ref{fig1}(a). Note
that it is symmetric about the central circle with $r=r_{0}$, i.e., $V(r=r_{0}-r^{\prime })=V(r=r_{0}+r^{\prime }$, for all $0<r^{\prime
}<r_{0}$, see Fig.~\ref{fig1}(b). This feature is important for the
stabilization of nonlinear modes due to the fact that the fields propagating
in the inner and outer annuli ($r<r_{0}$ and $r>r_{0}$, respectively)
experience the same guidance, in contrast to the case of the
harmonic-oscillator trapping potential, which is commonly used to guide
optical beams or trap BEC. Potential (\ref{V}) indeed favors the existence
of stable vortex solitons with high topological charges \cite{OL2023}, and
of stable higher-order solitons with a multiring profile, as shown by
additional numerical results.

Before unveiling properties of multipole modes produced by the full
nonlinear equation (\ref{Eq1}), it is instructive to understand the
dispersion relation of its linearized counterpart, as linear eigenvalues and
eigenmodes essentially affect the build of the nonlinear modes. The linear
spectrum produced by the numerical solution of the linearized version of
Eq.~(\ref{Eq2}), which is displayed in Fig.~\ref{fig1}(c), includes a finite
number of discrete real eigenvalues, in addition to the obvious continuous
spectrum (not shown here). The growth of the potential depth $p$ results in
a shift of the spectrum to the right. The variation of thickness $d$ and
radius $r_{0}$ also alters the distribution of the eigenvalues. Unlike the
system with the harmonic-oscillator potential, the discrete part of the
present spectrum is not equidistant. While the lowest (ground-state)
eigenvalue is nondegenerate (as it must be, according to the general
principles of quantum mechanics), the eigenvalues of the excited states are
doubly degenerate, with mutually orthogonal eigenmodes corresponding to
pairs of equal eigenvalues. Due to the degeneracy, the eigenmodes with $%
n\geq 2$ poles correspond to the $(2n-2)$-th and $(2n-1)$-th eigenvalues,
while the first eigenvalue corresponds to the ground-state eigenmode shown
in Fig.~\ref{fig1}(d). A linear superposition of the two degenerate
eigenmodes also produces an allowed state of the linear system. For example,
two superpositions of two orthogonal dipole modes (i.e., those oriented
along $x$ and $y$ directions), with $n=2$, build two counter-rotating vortex
states, which carry the angular momentum and opposite winding numbers
(topological charges) $m=\pm 1$.

\section{Numerical results for static multipole solitons}

\label{Sec3}

\begin{figure}[tbph]
\centering
\includegraphics[width=0.45\textwidth]{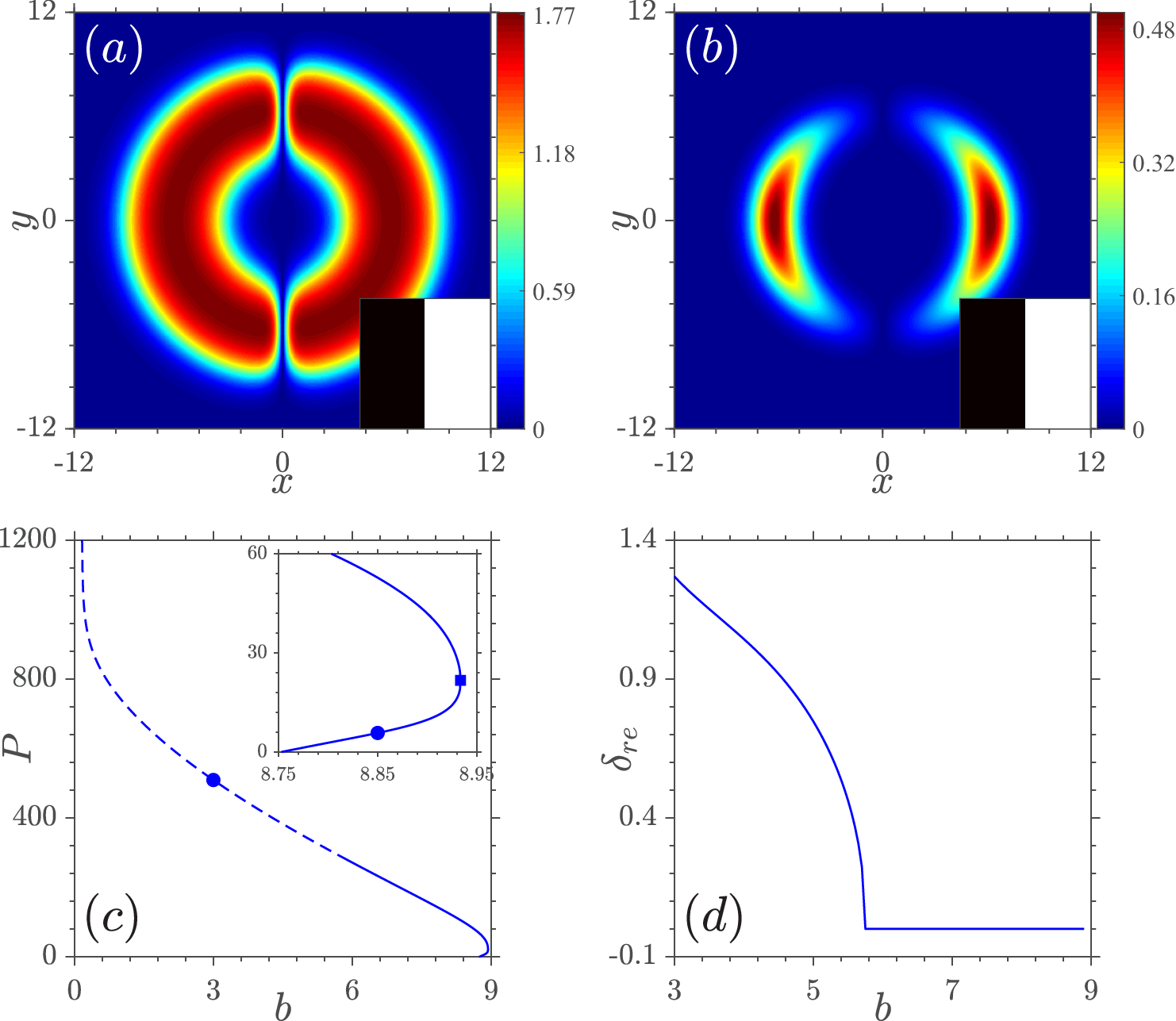}\vskip-0pc
\caption{(a) and (b): $\left\vert \protect\psi (x,y)\right\vert $ for the
unstable upper-branch and stable lower-branch dipole solitons with $b=3.0$
and $b=8.85$, which are marked by circles in the main plot and inset in (c),
respectively. Insets in (a) and (b) show the corresponding simple phase
distribution in the dipole modes. (c) Power of the dipole solitons $P$
versus propagation constant $b$. Inset: zoom of the $P(b)$ curve near the
turning point. The square denotes the merging point (cutoff of $b$). The solid and dashed lines denote, severally, stable and unstable segments of the dipole-soliton families. (d) The instability growth rate $\protect\delta _{\text{re}}$ versus $b$ for the upper-branch dipole
solitons. Parameters are $p=10,r_{0}=2\protect\pi $, $d=\protect\sqrt{6}$ in
all the panels.}
\label{fig2}
\end{figure}

Bearing the linear case in mind, we now focus on the properties of
multipole-mode solitons. First, dipole solitons bifurcate from the linear
eigenstates corresponding to the coinciding (mutually degenerate) second and
third eigenvalues, $b=8.753$, which are shown in Fig.~\ref{fig1}(c). The
power and amplitude initially increase with the growth of propagation
constant $b$, see the short bottom branch in the inset to Fig.~\ref{fig2}%
(c). As $b$ increases to $8.903$, the peak value of the dipole soliton
attains value $|\psi |_{\text{max}}=1$ (in the scaled notation adopted
here). Thus, the nonlinearity experienced by the dipole solitons is entirely
focusing at $b<8.903$. At $b>8.903$ ($|\psi |_{\text{max}}>1$), the effect
of the quintic defocusing becomes dominant in the core region with $|\psi
\left( x,y\right) |>1$. The defocusing nonlinearity in the core region
arrests the growth of $|\psi |_{\text{max}}$ and accelerates the increase of
the soliton's width, defined as
\begin{equation}
W^{2}\equiv \int \int (x^{2}+y^{2})\psi ^{2}\text{d}x\text{d}y/P.  \label{W}
\end{equation}%
The gradual transition to the defocusing nonlinearity, along with the action
of the trapping potential, prevents the existence of dipole solitons for the
propagation constant exceeding a cutoff value, \textit{viz}., at $b>b_{\text{%
cut}}=8.933$, the corresponding power being $P=21.693$. Instead, further
increase of $P$ is accommodated by the top branch of the $P(b)$ dependence
in Fig.~\ref{fig2}(c), with the negative slope, $dP/db<0$, which continues
the short bottom branch through the turning point, $b>b_{\text{cut}}$.

Dipole solitons belonging to the lower branch in Fig.~\ref{fig2}(c) are
composed of two far-separated crescent-shaped lobes with opposite signs,
tightly attached to the potential ring [Fig.~\ref{fig2}(b)]. With the
increase of $b$, the crescent lobes elongate, becoming thinner, while their
amplitude grows under the action of the dominant cubic focusing
nonlinearity. On the other hand, the dipole solitons belonging to the upper
branch in Fig.~\ref{fig2}(c) are shaped mainly by the dominant defocusing
quintic nonlinearity, which results in the expansion of the lobes along the
radial and azimuthal directions. As a result, the two lobes with opposite
signs are separated by narrow fissures, rather than wide gaps, see Fig.~\ref%
{fig2}(b). The expansion is more salient for solitons with higher power,
which tend to develop a flat-top profile.

The stability of the stationary states can be analyzed by taking perturbed
solutions to Eq.~(\ref{Eq1}) as
\begin{gather}
\Psi (x,y,z)=\left[ \psi (x,y)+f(x,y)\exp (\delta z)\right.  \notag \\
\left. +g^{\ast }(x,y)\exp (\delta ^{\ast }z)\right] \exp (ibz),
\label{pert}
\end{gather}%
where $f$ and $g$ are infinitesimal perturbations, and $\delta $ is the
instability growth rate. The linearization of Eq.~(\ref{Eq1}) around $\psi $
leads to an eigenvalue problem for $f(x,y)$ and $g(x,y)$:
\begin{equation}
i%
\begin{bmatrix}
\mathcal{M}_{1} & \mathcal{M}_{2} &  \\
-\mathcal{M}_{2}^{\ast } & -\mathcal{M}_{1}^{\ast } &
\end{bmatrix}%
\begin{bmatrix}
f \\
g \\
\end{bmatrix}%
=\delta
\begin{bmatrix}
f \\
g \\
\end{bmatrix}%
.  \label{Eq4}
\end{equation}%
Here, ${\mathcal{M}_{1}}\equiv \frac{\partial ^{2}}{\partial x^{2}}+\frac{%
\partial ^{2}}{\partial y^{2}}+pV-b+2\left\vert \psi \right\vert ^{2}-3|\psi
|^{4}$, ${\mathcal{M}_{2}}\equiv \psi ^{2}(1-2|\psi |^{2})$, and $\ast $
denotes the complex conjugate. Equations (\ref{Eq4}) can be solved by means
of the Fourier collocation method \cite{yang20103}. Solitons are stable if
all eigenvalues $\delta $ are imaginary.

Instability growth rates for the upper- and lower-branch dipole solitons
were obtained from the numerical solutions of Eqs.~(\ref{Eq4}). The results
demonstrate that the lower branch is stable in its entire existence domain,
while the upper branch is stable if the propagation constant exceeds a
certain critical value (which is a generic feature of 2D NLSEs with the CQ
nonlinearity), \textit{viz}., $b>b_{\text{cr}}=5.751$ [see Fig.~\ref{fig2}%
(c)]. At $b<b_{\text{cr}}$, the instability growth rate rapidly increases
with the decrease of $b$, see Fig.~\ref{fig2}(d).

\begin{figure}[tbph]
\centering
\includegraphics[width=0.48\textwidth]{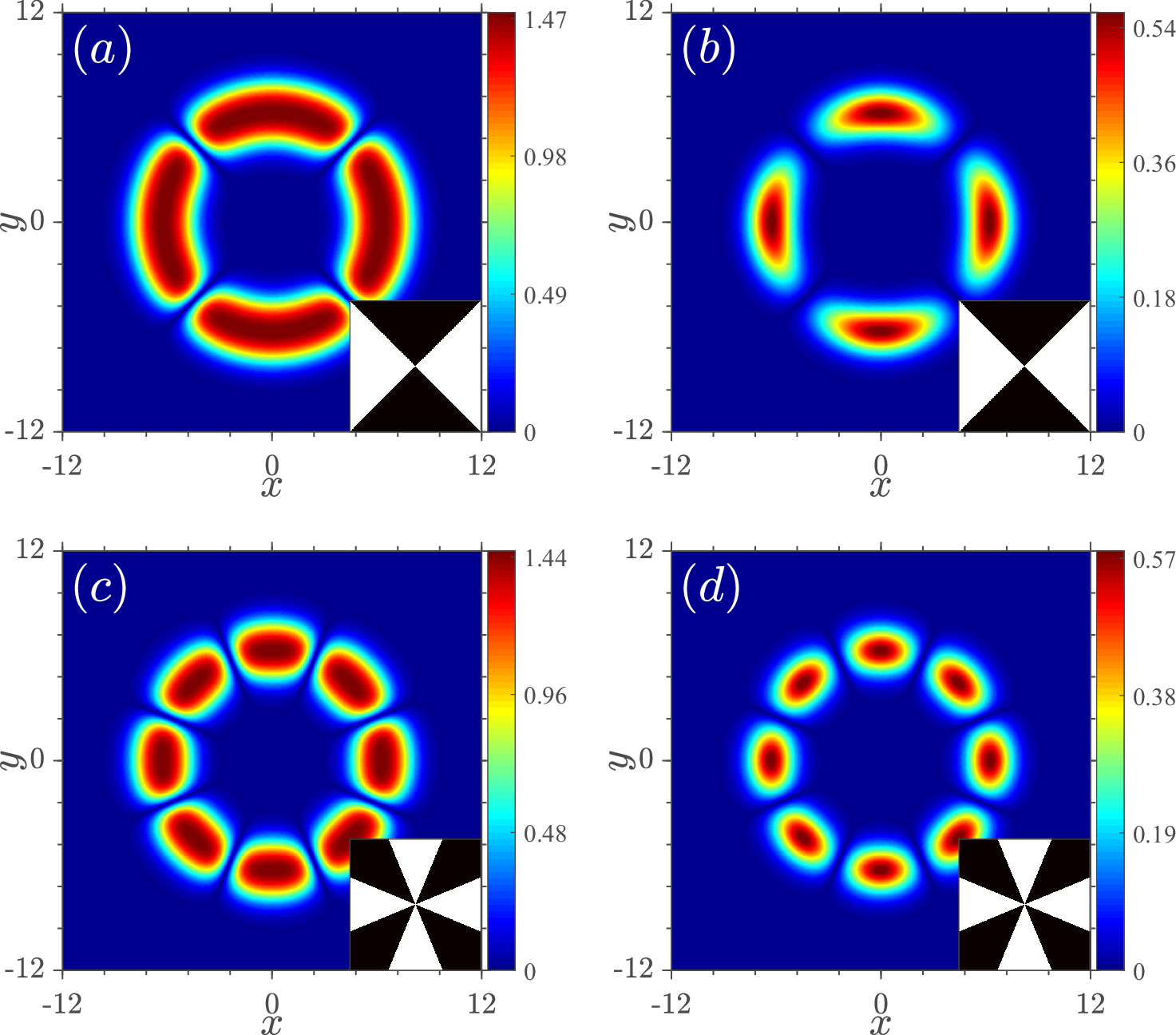}\vskip-0pc
\caption{(a) and (b): $\left\vert \protect\psi \left( x,y\right) \right\vert
$ for stable upper-branch and unstable lower-branch quadrupole solitons with
$b=7.0$ and $b=8.8$, respectively, which are marked in Fig.~\protect\ref%
{fig4}(a). (c) and (d): The same for upper- and lower-branch stable and
unstable octupole solitons with $b=7.2$ and $b=8.5$, respectively, which are
marked in Fig.~\protect\ref{fig4}(b). }
\label{fig3}
\end{figure}

Next, we address multipole-mode solitons with a larger number $n$ of the
lobes. The patterns of $\left\vert \psi \left( x,y\right) \right\vert $ for
quadrupole and octupole solitons are shown in Fig.~\ref{fig3}. Similar to
the dipoles, the lobes forming the lower-branch quadrupoles are separated by
relatively wide fissures [Fig.~\ref{fig3}(b)], while the lobes of their
upper-branch counterparts are \textquotedblleft fatter", being tightly
packed along the potential ring [see Fig.~\ref{fig3}(a)]. These features
indicate that the repulsion between adjacent lobes of the upper-branch
quadrupoles is strong.

Adjacent lobes of octupole solitons are pressed onto each other tighter than
in the quadrupoles, cf. Figs.~\ref{fig3}(b) and \ref{fig3}(c). Due to the
confinement imposed by the annular (ring-shaped) potential, the radial size
of the multipole solitons (the distance from the center of each lobe to the
origin) does not change with the growth of $n$. As mentioned above, the
adjacent lobes in all multipole solitons have alternating signs, see inset
phase plots in Fig.~\ref{fig3}.

\begin{figure}[tbph]
\centering
\includegraphics[width=0.48\textwidth]{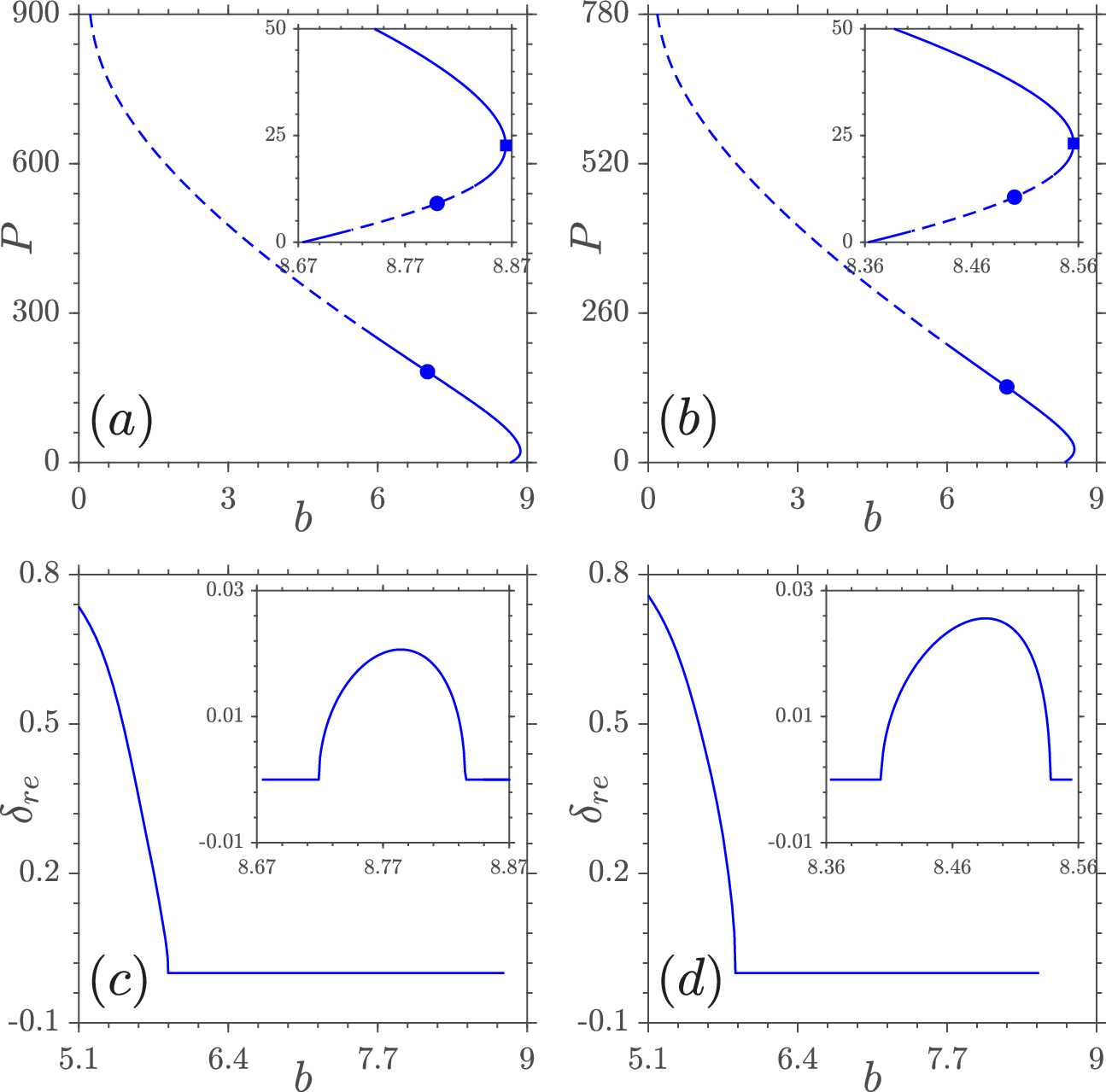}
\caption{(a) and (b): Integral power $P$ versus propagation constant $b$ for
quadrupole and octupole solitons, respectively. Insets: the $P(b)$ curves
near the merging (turning) point. Solid and dashed lines denote stable and
unstable solitons. (c) and (d): The instability growth rate $\protect\delta %
_{\text{re}}$ versus $b$ for the upper-branch and lower-branch (inset)
quadrupole and octupole solitons, respectively.}
\label{fig4}
\end{figure}

The lower $P(b)$ branch for the quadrupole solitons, originating from the
linear mode at $b=8.674$, merges with the upper branch at $b_{\text{cut}%
}=8.864$ [see Fig.~\ref{fig4}(a)]. The merging (turning) point for the
octupole solitons is $b_{\text{cut}}=8.555$, see Fig.~\ref{fig4}(b). The
shift of $b_{\text{cut}}$ for solitons with different numbers of poles is a
straightforward consequence of the difference in the linear eigenvalues of $%
b $ from which the lower-branch solitons bifurcate. The upper $P(b)$ curves can extend to still larger values of $P$ for smaller $b$. We do not pay more attention to the corresponding multipole solitons as they are definitely unstable.

Unlike the dipole solitons which are completely stable on the lower branch
[see Fig.~\ref{fig2}(c)], weak instability occurs for the lower-branch
quadrupole and octupole solitons, as seen in insets in Figs.~\ref{fig4}(a)
and (b). The instability region expands slowly with the growth of the pole
number, as seen in insets to Figs.~\ref{fig4}(c), \ref{fig4}(d) and \ref%
{fig5}(d). Further, the upper-branch quadrupole and octupole solitons are
stable, severally, in intervals $b\in \lbrack 6.013,8.864]$ and $b\in
\lbrack 5.859,8.555]$. as shown in Figs.~\ref{fig4}(c) and (d). With the
increase of $n$, the stability segment of the upper-branch multipole
solitons shrinks \emph{very slowly}, which is the most important finding of
this work. For example, the width of the stability intervals for the
quadrupoles and octupoles are $2.851$ and $2.696$, respectively. Actually,
these intervals, located to the left of the merging point $b_{\text{cut}}$,
are relatively large ones. We stress that the stability of the multipole
solitons trapped in the annular potential is very different from that for
vortex solitons trapped in the same potential \cite{OL2023}, where the
lower-branch vortices with all topological charges are unstable in their
almost entire existence domain, while all upper-branch vortices are
completely stable.

\begin{figure}[tbph]
\centering
\includegraphics[width=0.48\textwidth]{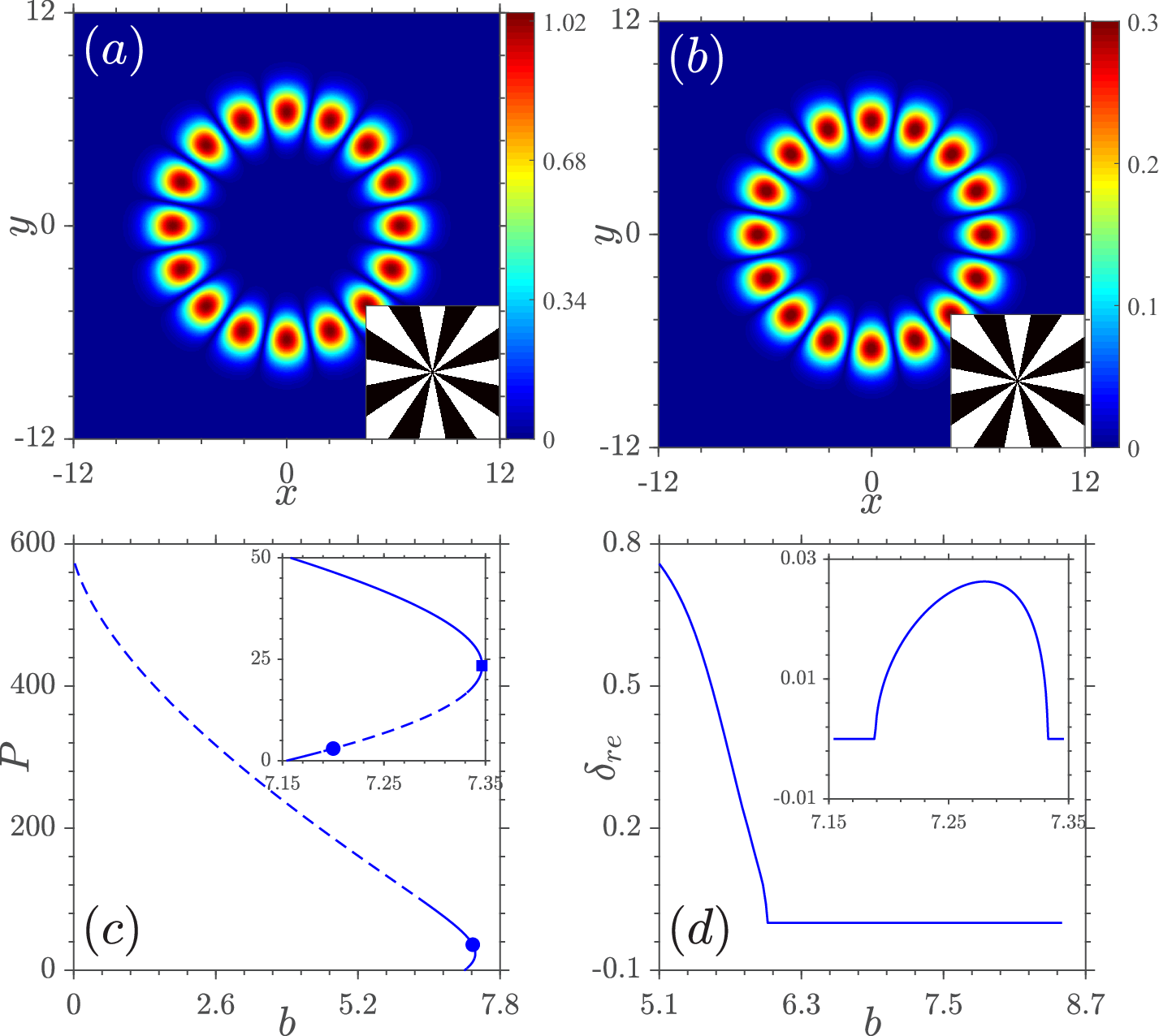}
\caption{(a) and (b): $\left\vert \protect\psi \left( x,y\right) \right\vert
$ of $16$-pole solitons with $b=7.3$ and $7.2$, which are marked,
respectively, in the main plot and inset in (c). Insets in (a) and (b) show
the corresponding phase distribution. (c) The $P(b)$ dependence for the $16$%
-pole solitons. Inset: The $P(b)$ curve near the merger (turning) point. The
solid and dashed lines denote stable and unstable segments of the soliton
families, respectively. (d) The instability growth rate $\protect\delta _{%
\text{re}}$ versus $b$ for the upper branch and lower branch (inset) of the $%
16$-dipole soliton family.}
\label{fig5}
\end{figure}

To further understand the properties of multipole solitons, we investigated
necklace-shaped ones with $n=10,12,14$, and $16$. Representative examples of
the necklaces with $n=16$ poles are presented in Figs.~\ref{fig5}(a) and
(b). Due to the confinement imposed by the annular potential, the components
in both lower- and upper-branch solitons are tightly packed in the azimuthal
direction and somewhat stretched along the radius. This effect is more
evident for the solitons with a higher power or higher $n$. The lower-branch
necklace solitons with $n=16$ bifurcate from the linear mode at $b=7.154$,
while the point of the merger of the lower and upper branches is at $b_{%
\mathrm{cut}}=7.346$. The stronger repulsion between the adjacent lobes in
the case of larger $n$ causes shrinkage of the stability interval for the
upper-branch solitons. Nevertheless, for $n=16$ there still exists a
sufficiently wide stability interval to the left of $b_{\text{cut}}$, see
Fig.~\ref{fig5}(d). A nonvanishing stability region can be found for still
higher values of $n$ (e.g., $n=20$). These findings are in sharp contrast to
the in 2D cubic-quintic NLSE\ with the harmonic-Gaussian potential \cite%
{liu20232}, where the stability region is virtually invisible for $n>8$.

\begin{figure}[tbph]
\centering
\includegraphics[width=0.48\textwidth]{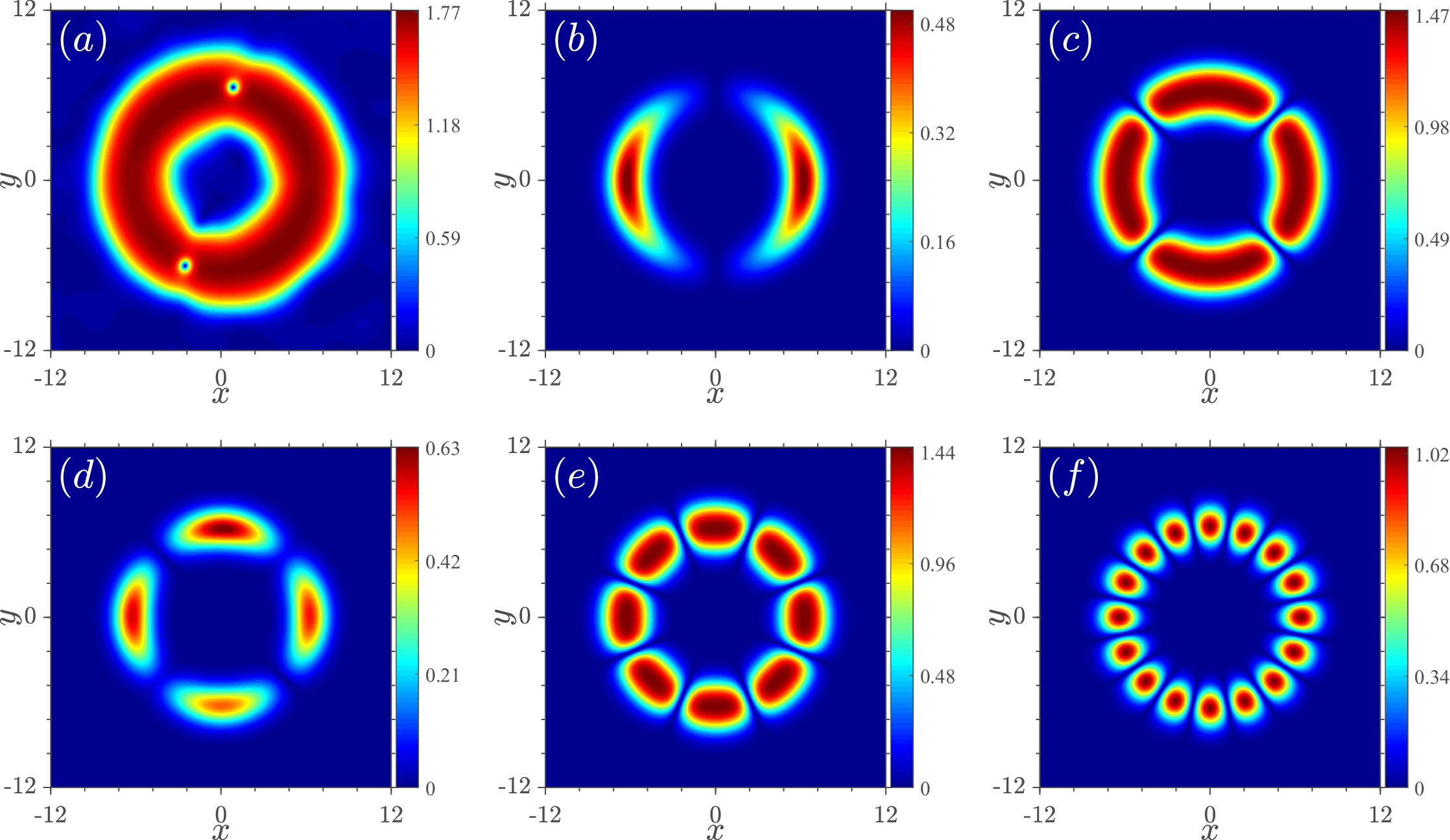}
\caption{Examples of the simulated propagation of unstable (a,d) and stable
(b,c,d,f) solitons with $n=2$ (a,b), $n=4$ (c,d), $n=8$ (e), and $n=16$ (f).
The solitons belong to the lower branch in (b,d), and to the upper branch in
other panels. The propagation constant is $b=3.0$ in (a), $8.85$ in (b), $%
7.0 $ in (c), $8.8$ in (d), $7.2$ in (e), and $7.3$ in (f). The panels
exhibit the outcome of the propagation after passing the distance $z=100$ in
(a), $1540$ in (d) and $2000$ in other panels.}
\label{fig6}
\end{figure}

To verify the predictions of the linear-stability analysis for the multipole
families with different values of $n$, we conducted extensive simulations of
perturbed propagation for the solitons, by means of the split-step Fourier
method. The perturbation was added, as white noise, to the input at $z=0$
for stable solitons, and no perturbations were added to the solitons whose
instability was predicted by the computation of eigenvalues for small
perturbations. Typical examples of stable and unstable propagation are
presented in Fig.~\ref{fig6}. For the upper-branch dipole soliton with $%
b=3.0 $, the large instability growth rate $\delta _{\text{re}}=1.266$
destroys the original pattern, shown in Fig.~\ref{fig2}(a), after a short
propagation distance, see Fig.~\ref{fig6}(a). The lower-branch quadrupole
soliton with $b=8.8$ and a small instability growth rate, $\delta _{\text{re}%
}=0.019$, is only deformed, but not destroyed, after a long distance [Fig.~%
\ref{fig6}(d)]. White noise added to the stable solitons is quickly radiated
away in Figs.~\ref{fig6}(b,d-f). The agreement between the predictions of
the linear-stability analysis and direct simulations is obtained for all the
cases considered.

\section{Rotating multipole solitons: analytical and numerical results}

Because the multipole solitons have the articulate structure in the
azimuthal direction, it is natural to try setting them in rotation, by
imprinting a phase torque,
\begin{equation}
\psi \left( x,y\right) \rightarrow \psi \left( x,y\right) \exp (im\theta ),
\label{m}
\end{equation}%
with integer index $m$, onto the static soliton. First, the result of the
application of this \textquotedblleft crank" to the static multipole with
the narrow radial shape can be predicted in an approximate analytical form.
Indeed, assuming that the result is a pattern rotating with angular velocity
$\Omega $, in the form of
\begin{equation}
\Psi \left( x,y,z\right) =\varphi \left( r,\theta -\Omega z\right) \exp
\left( ibz+im\theta \right)   \label{rot}
\end{equation}%
[cf. Eq.~(\ref{Psipsi})], which is written in the polar coordinates $\left(
r,\theta \right) $, we substitute ansatz (\ref{rot}) in the underlying
equation (\ref{Eq1}). This\ leads to the conclusion that wave function $%
\varphi $ satisfies the stationary equation%
\begin{equation}
b\varphi =\frac{\partial ^{2}\varphi }{\partial r^{2}}+\frac{1}{r}\frac{%
\partial \varphi }{\partial r}+\frac{1}{r^{2}}\frac{\partial ^{2}\varphi }{%
\partial \tilde{\theta}^{2}}+\left[ pV(r)-\frac{m^{2}}{r^{2}}\right] \varphi
+|\varphi |^{2}\varphi -|\varphi |^{4}\varphi   \label{varphi}
\end{equation}%
[cf. Eq.~(\ref{Eq2})], where $\tilde{\theta}\equiv \theta -\Omega z$, the
term $-m^{2}/r^{2}$ representing the centrifugal energy. An additional
relation produced by the substitution of ansatz (\ref{rot}) in Eq. (\ref{Eq1}%
) is an \emph{approximate} one, $\Omega =2m/r^{2}$. It is approximate
because it produces $\Omega $ as a function of $r$, while the angular
velocity must be a constant. However, for the pattern confined to a narrow
annulus around $r=r_{0}$, one may, in the first approximation, replace $r$
by $r_{0}$, thus predicting the angular velocity and respective rotation
period,%
\begin{equation}
\Omega \approx 2m/r_{0}^{2},~Z=2\pi /\Omega \approx \pi m^{-1}r_{0}^{2}.
\label{angular}
\end{equation}%
In particular, for the radial size $r_{0}=2\pi $ of the trap adopted here
[see Eq.~(\ref{r0})] and $m=1$, Eq.~(\ref{angular}), predicts the rotation
period $Z\approx $ $4\pi ^{3}\approx \allowbreak 124$, while the value
produced by direct simulations is $Z\approx 120$ (see below), hence the
approximation is quite accurate. Furthermore, the simulation for the torque
with $m=2$ produces the rotation with the double angular velocity, in
agreement with prediction given by Eq. (\ref{angular}).

The competition of the centrifugal energy with the trapping potential in
Eq.~(\ref{varphi}) may eventually suppress the trapping effect if the
effective combined potential, $-pV(r)+m^{2}/r^{2}$, has no minimum. In the
lowest approximation, the condition for the survival of the minimum is%
\begin{equation}
r_{0}^{3}>\sqrt{2e}m^{2}d/p.  \label{r0^3}
\end{equation}

\begin{figure}[tbph]
\centering
\includegraphics[width=0.48\textwidth]{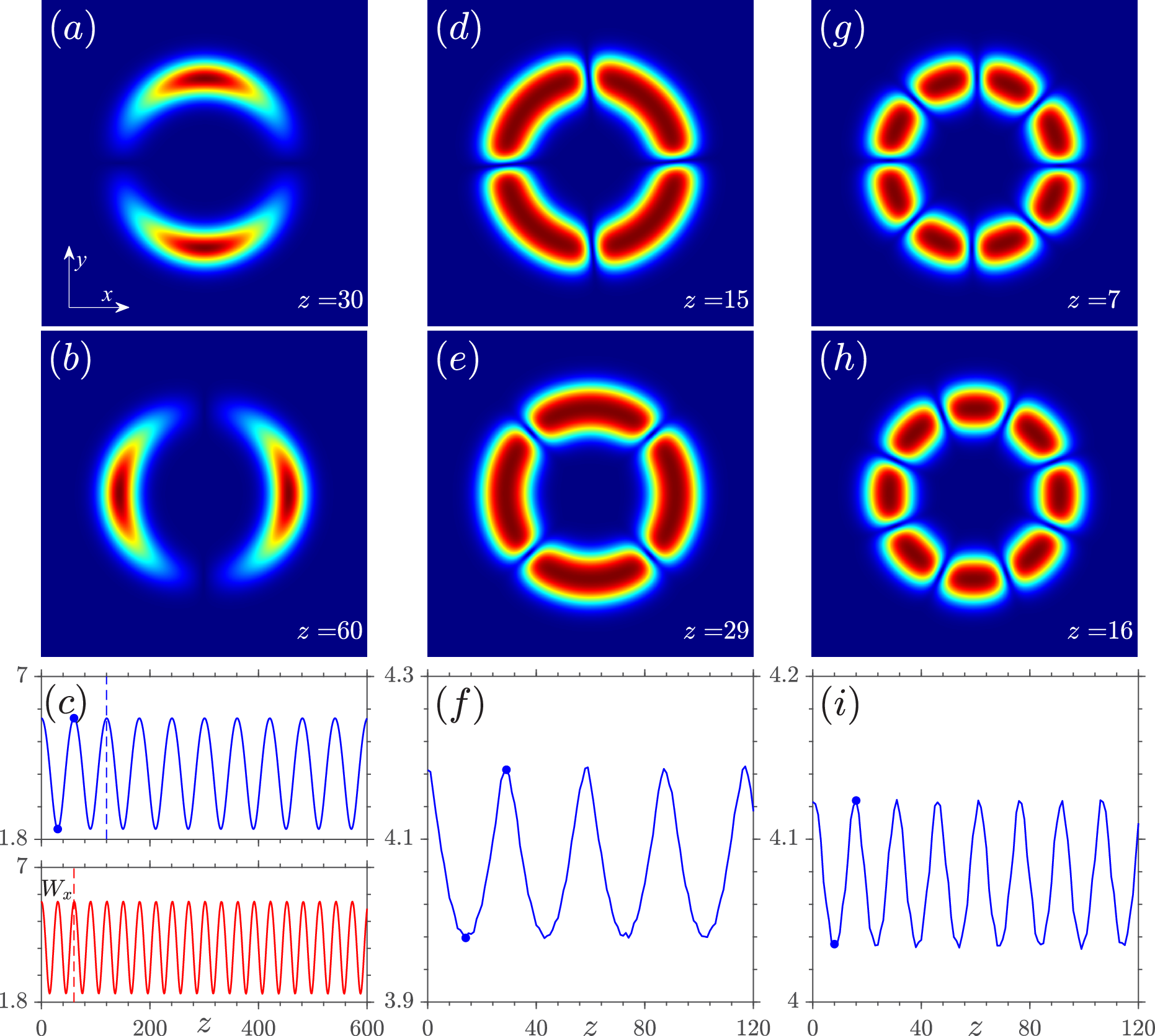}
\caption{The stable counterclockwise rotation of dipole (a-c), quadrupole
(d-f), and octupole (g-i) solitons initiated by torque (\protect\ref{m})
with $m=1$, applied to inputs selected as the stationary solitons presented
in Figs.~\protect\ref{fig6}(b), (c), and (e), respectively. The periods of
oscillations of the $x$-width observed in panels (c), (f), and (i) agree
with the prediction $Z/n$, see the main text. The lower plot in (c) displays
the oscillations of the $x$-width of the same dipole soliton initiated by
the double torque, with $m=2$. In panel (c), the vertical dashed lines
define one rotation period, $Z$.}
\label{fig7}
\end{figure}

Systematic simulations clearly confirm transition to a regime of steady
rotation, initiated by the application of torque (\ref{m}). The rotation
lasts for indefinitely many periods, without generating any tangible loss
[see Fig.~\ref{fig7}]. In particular, the torque with $m=1$ gives rise to
the robust rotational state with period $Z\approx 120$.

The rotation is coupled to periodic oscillations of the effective soliton's
width. In particular, the oscillations of the width in the $x$-direction,
defined by%
\begin{equation}
W_{x}^{2}\equiv \int \int x^{2}\psi ^{2}\text{d}x\text{d}y/P\approx W^{2}/2
\label{Wx}
\end{equation}%
[cf. Eq.~(\ref{W})], with period $Z\approx 120$, initiated by the torque
with $m=1$ in Eq.~(\ref{m}), are plotted in Figs.~\ref{fig7}(c,f), and (i)
for the rotating dipole, quadrupole, and octupole, respectively. Note that
the period of oscillations of width (\ref{Wx}) exhibited by a $n$-multipole
rotating with period $Z$ is $Z/n$. The oscillations observed in Figs.~\ref{fig7}(c,f,i) completely agree with this expectation.

To illustrate the steady rotation, examples of the patterns are
displayed at $z=Z/4$ and $Z/2$ for the dipole soliton [Figs.~\ref{fig7}%
(a,b)], $Z/8$ and $Z/4$ for the quadrupole [Figs.~\ref{fig7}(d,e)], and $Z/16
$ and $Z/8$ for the octupole [Figs.~\ref{fig7}(g,h)]. The robustness of the
rotation is corroborated, in particular, by Fig.~\ref{fig7}(c), which shows
an example of long evolution, for $z=600$ (ca. $5$ full rotation periods).
Movies presented in the Supplemental material \cite{SupplMovie} illustrate
the rotation dynamics in full detail (the third movie displays the clockwise
rotation, while others represent the opposite direction). Robust rotary
states of the multispot patterns suggest new possibilities for all-optical
routing of weak data-carrying light beams \cite{kartashov2005}.

The multipole solitons predicted in this work can be created in currently
available optical experimental setups. In particular, polydiacetylene
paratoluene sulfonate (PPS) exhibits competing nonlinearities with
sufficiently large cubic and quintic indices \cite{Lawrence:98}. For a beam
with carrier wavelength $\lambda =1.6$ $\mathrm{\mu }$m, the second- and
fourth-order optical indices are $n_{2}=2.2\times 10^{-3}\text{cm}^{2}/\text{%
GW}$ and $n_{4}=0.8\times 10^{-3}\text{cm}^{4}/\text{GW}^{2}$, respectively.
The critical intensity at which $\delta n=0$ is $I_{0}=|n_{2}/n_{4}|=2.75%
\text{ GW/}\text{cm}^{2}$. This material exhibits focusing nonlinearity at $%
I<0.5I_{0}$ and becomes defocusing at $I>0.5I_{0}$. In other words, a
localized beam experiences self-defocusing around the peak intensity
(assuming $I_{\text{max}}>0.5I_{0}$) and focusing in the wings, where $%
I<0.5I_{0}$. As concerns the linear refractive-index modulation inducing the
annular trapping potential, it may be created by means of the
well-elaborated technology used for the production of multi-layer fibers
\cite{multi-layer}.

\section{Conclusion}

\label{Sec4} Summarizing, we have investigated the existence, stability, and
propagation dynamics of excited nonlinear states containing different even
numbers $n$ of lobes in CQ (cubic-quintic) optical media equipped with an
annular (ring-shaped) trapping potential. Adjacent components of such
solitons have opposite signs, being uniformly placed along the potential
annulus. With the increase of the soliton's amplitude, the transition from
focusing to defocusing nonlinearity, along with the action of the trapping
potential, gives rise to two branches of multipole soliton families, with
opposite slopes of the power-vs.-propagation-constant curves. With the
increase of $n$, the stability domain of the solitons shrinks very slowly,
allowing the tangible presence of stable solitons with $n>16$. Steady
rotation of multipole solitons, initiated by the application of the phase
torque, is predicted analytically and numerically. The results suggest new
ways of manipulating light signals and the creation of novel self-trapped
states. The analysis reported here can be generalized for optical solitons
in competing quadratic-cubic media, matter-wave solitons in Bose-Einstein
condensates, and quantum droplets in Bose-Bose mixtures trapped in a ring
potential. A challenging possibility is to extend the consideration for
elliptically-shaped trapping potentials.

\vskip0.5pc

\vskip0.5pc \textbf{Acknowledgements:} This work is supported by the Natural
Science Basic Research Program in Shaanxi Province of China (Grant No.
2022JZ-02) and Israel Science Foundation (Grant No. 1695/22).


%
%
%


\end{document}